\documentclass[iop]{emulateapj} 
\usepackage[dvipsnames]{xcolor}  
\usepackage{amsmath}  
\usepackage[colorlinks=true,
            breaklinks=true,
            linkcolor=Red,
            citecolor=MidnightBlue,
            urlcolor=MidnightBlue]{hyperref}
\usepackage[T1]{fontenc}  

\newcommand\ergs{erg~s$^{-1}$}

\newcommand{\eddr}{$\lambda_{\rm Edd}$}

\newcommand{\nhgal}{$N_{\rm H0}$}
\newcommand{\nhint}{$N_{\rm H}$}
\newcommand{\nhtot}{$N_{\rm H1}$}

\newcommand{\GalNum}{719}        

\newcommand{\chandra}[0]{\textit{Chandra}}

\begin{document}
\title{{\it Chandra}\ Survey of Nearby Galaxies: Testing the Accretion Model for Low-luminosity AGNs}

\author{Rui She\altaffilmark{1},
Luis C.\ Ho\altaffilmark{2,3},
Hua Feng\altaffilmark{1}, and
Can Cui\altaffilmark{4}}
\altaffiltext{1}{Department of Engineering Physics and Center for Astrophysics, Tsinghua University, Beijing 100084, China}
\altaffiltext{2}{Kavli Institute for Astronomy and Astrophysics, Peking University, Beijing 100087, China}
\altaffiltext{3}{Department of Astronomy, Peking University, Beijing 100087, China}
\altaffiltext{4}{Shanghai Astronomical Observatory, Chinese Academy of Sciences, 80 Nandan Road, Shanghai 200030, China}

\tabletypesize{\scriptsize}

\begin{abstract}

  From a {\it Chandra} sample of active galactic nuclei (AGNs) in nearby galaxies, we find that for low-luminosity AGNs (LLAGNs), either the intrinsic absorption column density, or the fraction of absorbed AGNs, positively scales with the Eddington ratio for $L_{\rm bol}/L_{\rm Edd} \lesssim 10^{-2}$.  Such a behavior, along with the softness of the X-ray spectrum at low luminosities, is in good agreement with the picture that they are powered by hot accretion flows surrounding supermassive black holes.  Numerical simulations find that outflows are inevitable with hot accretion flows, and the outflow rate is correlated with the innermost accretion rate in the low-luminosity regime. This agrees well with our results, suggesting that the X-ray absorption originates from or is associated with the outflow material. Gas and dust on larger scales may also produce the observed correlation.  Future correlation analysis may help differentiate the two scenarios.
 
\end{abstract}


\section{Introduction}
\label{sec:intro}

Accreting supermassive black holes are the central engine of active galactic nuclei (AGNs) \citep{antonucci93,ho08}. Understanding the accretion physics for supermassive black holes is important not only in its own right, but also for understanding the impact of AGN feedback on the environment via radiation and outflows \citep{Fabian2012,kormendy13}, which is an essential ingredient in the evolution of galaxies and large-scale structures.

The details of the accretion process may vary dramatically as a function of the accretion rate. The standard accretion disk model \citep{shakura73,Koratkar_Blaes_1999} can successfully explain the AGN power in the regime of moderate accretion rates (say, $\sim 10^{-2}$--$10^{-1}$ of the Eddington accretion rate, $\dot{M}_{\rm Edd} = L_{\rm Edd}/0.1c^2$), where a radiative efficiency of typically 0.1 is expected.  The standard disk is geometrically thin and optically thick, and viscous dissipation balances radiative cooling locally. At higher accretion rates near or around the Eddington limit, (e.g., for narrow-line Seyferts 1s), the slim disk model \citep{Abramowicz_etal_1988} is invoked instead, where, due to a large photon diffusion timescale, the radiation may be trapped in the disk and advected onto the central black hole, leading to a lower radiative efficiency than that for the standard disk model. 

Toward the low-accretion regime (e.g., $\lesssim 10^{-3} \dot{M}_{\rm Edd}$), which is the topic of interest of this paper, the standard cool disk model is no longer able to fit the observations. These low-luminosity AGNs (LLAGNs) are found to be radiatively inefficient rather than starved \citep{ho08,ho09,pellegrini05,soria06c,paggi16}. Besides, the absence of both the Fe K$\alpha$ emission line \citep{fabbiano03,ptak04,binder09,younes11,kawamuro13} and the ``big blue bump'' \citep{ho99,ho08,chiaberge06,eracleous10a,younes12} also implies that they have an accretion geometry different from luminous AGNs.  In this regime, models with hot accretion flows are proposed \citep[for a recent review see][]{yuan14a}, among which the most widely studied category of models is called advection-dominated accretion flows \citep[ADAFs; e.g,][]{Narayan_Yi_1994}. It was proposed that these hot accretion flows are of low radiative efficiency due to several reasons: advection of the accretion power locked in the hot ions onto the black hole \citep{Narayan_Yi_1995}, convection \citep{Narayan_etal_2000}, and the presence of outflows \citep{Blandford_Begelman_1999}. Numerical simulations have shown that a strong wind does exist while convection is absent (\citealt{YuanBuWu2012,YuanWuBu2012,narayan12,yuan15}; see \citealt{yuan14a} for a review).

The central black hole Sgr~A$^\ast$ in our Milky Way, due to its vicinity, is a powerful laboratory to test hot accretion flow models. The measured radiative efficiency in its quiescent state is as low as $\sim$$10^{-6}$ \citep{Baganoff_etal_2003}. \citet{Yuan_etal_2003} suggested that, in addition to the traditional ADAF, outflows play an important role in reducing the mass accretion rate and radiative efficiency. This was indirectly confirmed with deep \textit{Chandra} observations \citep{Wang_etal_2013}.

Numerical simulations suggest that the outflow rate scales with the accretion rate in the LLAGN regime \citep{yuan15}; as the accretion rate is further decreased, so is the mass loss rate in the form of outflow. The outflow may contribute a significant fraction of the absorption column density in the X-ray band. \citet{beckmann09} found that the fraction of absorbed (intrinsic \nhint $> 10^{22} \; {\rm cm}^{-1}$) Seyfert galaxies declines when $L_{\rm 20-100~keV} < 3 \times 10^{41}$~\ergs, although there are only three AGNs in this luminosity bin. \citet{zhang09} found a trend of decreasing \nhint \footnote{In this paper, we use the notation \nhint\ for intrinsic absorption, \nhgal\ for Galactic absorption, and \nhtot\ for total absorption, i.e., \nhtot=\nhint+\nhgal .} with decreasing X-ray luminosity in the range of $10^{37}$--$10^{42}$~\ergs.  These results are generally consistent with the prediction, but more tests are needed.

Some studies also relate the outflow to the formation of the dust torus \citep{elitzur06}, suggesting that the broad-line region is located inside the dust sublimation radius. A natural consequence of such a scenario is that the dust torus and  broad-line region cannot be sustained at very low accretion rates \citep{elitzur09}.

Nearby LLAGNs are ideal targets to test these models. \citet[][hereafter Paper I]{she17} recently conducted an archival X-ray survey of nearby galaxies with \chandra/ACIS observations. In this brief paper, we show that LLAGNs do follow the prediction that their intrinsic absorption column density scales with the accretion rate, consistent with results from numerical simulations.  We also report the dependence of the power-law photon index with the accretion rate, and explain the results in the framework of the hot accretion flow model. 

\section{Sample, spectral fitting, and results}
\label{sec:data}

We constructed a sample of 314 AGN candidates out of 719 galaxies within 50 Mpc from the \chandra\ public archive as of March 2016 (see Paper I for more details). X-ray spectra were extracted for 154 sources that have more than 100 photons in the energy range 0.3--8 keV. The task \textit{specextract} was employed to extract the energy spectra, which were grouped to have at least 15 counts in each spectral bin. The spectra were fitted with physical models in XSPEC 12.8 \citep{arnaud96}. For each spectrum, we first tried a simple power-law model ({\tt powerlaw} in XSPEC) subject to interstellar absorption ({\tt wabs}), with one component fixed at the Galactic absorption column density \citep{kalberla05} and the other allowed to vary to account for extragalactic or intrinsic absorption. If the second component converges to zero, we then remove it and place an upper limit of $10^{20}$~cm$^{-2}$ for the intrinsic \nhint. If the single power-law model is inadequate to fit the data, resulting in a null hypothesis probability less than 0.05, we attempted to add a thermal plasma component ({\tt mekal}) or a cool blackbody component ({\tt bbody}) without intrinsic absorption, or replace the intrinsic absorption ({\tt wabs}) with a partial covering absorption ({\tt pcfabs}), or, occasionally, even more complex models (see Paper I for more details). The significance of the additional component is evaluated with the F-test; it is valid if the chance probability is $<0.05$.  In summary, for the best-fit models, 74 spectra can be fitted with the model {\tt wabs $\ast$ powerlaw}, 53 with {\tt wabs $\ast$ powerlaw + mekal}, 10 with {\tt pcfabs $\ast$ powerlaw}, 12 with {\tt pcfabs $\ast$ powerlaw + mekal}, 1 with {\tt pexmon}, 2 with {\tt wabs $\ast$ (powerlaw + bbody)}, and 2 with {\tt wabs $\ast$ (powerlaw + bbody) + mekal }.  

We note that some AGNs are embedded in a diffuse gas component that can be resolved on \chandra\ images. When we choose the background aperture, we do not attempt to remove the contribution from the diffuse gas component, as it may vary with location.  Instead, we model it, if needed, using a  {\tt mekal} component. Visual inspection of the X-ray images suggests a strong correlation between the presence of the {\tt mekal} component in the spectrum and diffuse gas on the image. As the AGN luminosity is quoted in the energy range of 2--10 keV, contamination from the {\tt mekal} component is negligible because its temperature is mostly in the range 0.4--0.9 keV \citep[][ Paper I]{ho08}.

To convert the luminosity to Eddington ratio (\eddr), the black hole mass is estimated using a re-calibrated $M_{\rm BH} - \sigma_\ast$ relation\footnote{We perform a linear regression between the black hole masses and bulge stellar velocity dispersions for all galaxies with reliable measurements of both quantities as given in \citet{kormendy13}.  Unlike \citet{kormendy13}, who only provided fits for classical bulges and elliptical galaxies, we do not distinguish between bulge types.  The resulting $M_{\rm BH} - \sigma_\ast$ relation is $\log [M_{\rm BH}/(10^9~M_{\sun})]=-(0.68\pm0.05)+(5.20\pm0.37)\log[\sigma/(200~{\rm km~s^{-1}})]$. See Paper I for details.}, where $\sigma_\ast$ is the central stellar velocity dispersion of the host galaxy. Although the extrapolation to galaxies with a pseudo-bulge or those without a bulge component may be problematic, this is the best one can do for the whole sample.  All the spectral quantities are listed in Paper I. 

\begin{figure}[htbp]
\centering
\includegraphics[width=0.77\columnwidth]{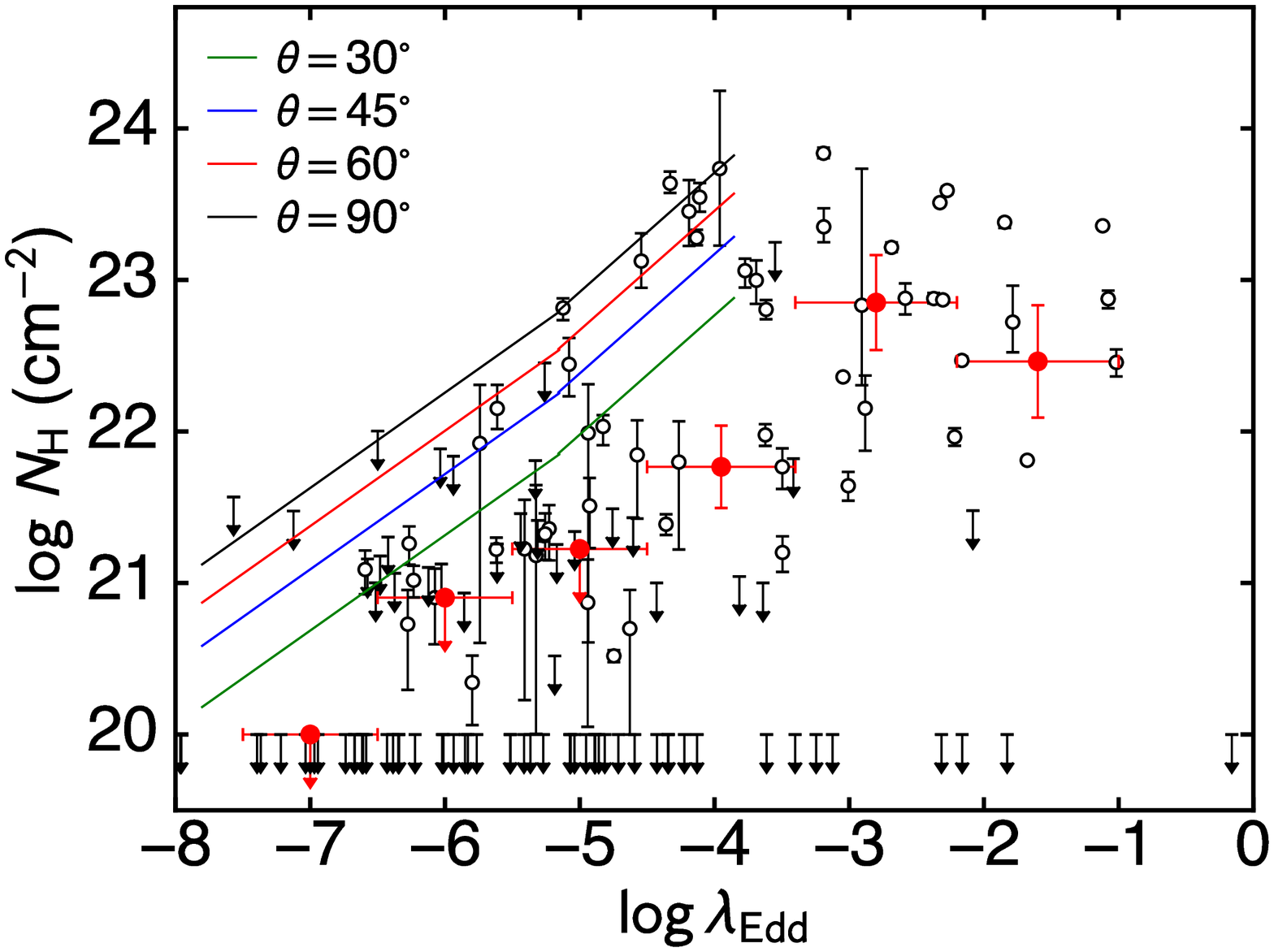}\\
\includegraphics[width=0.77\columnwidth]{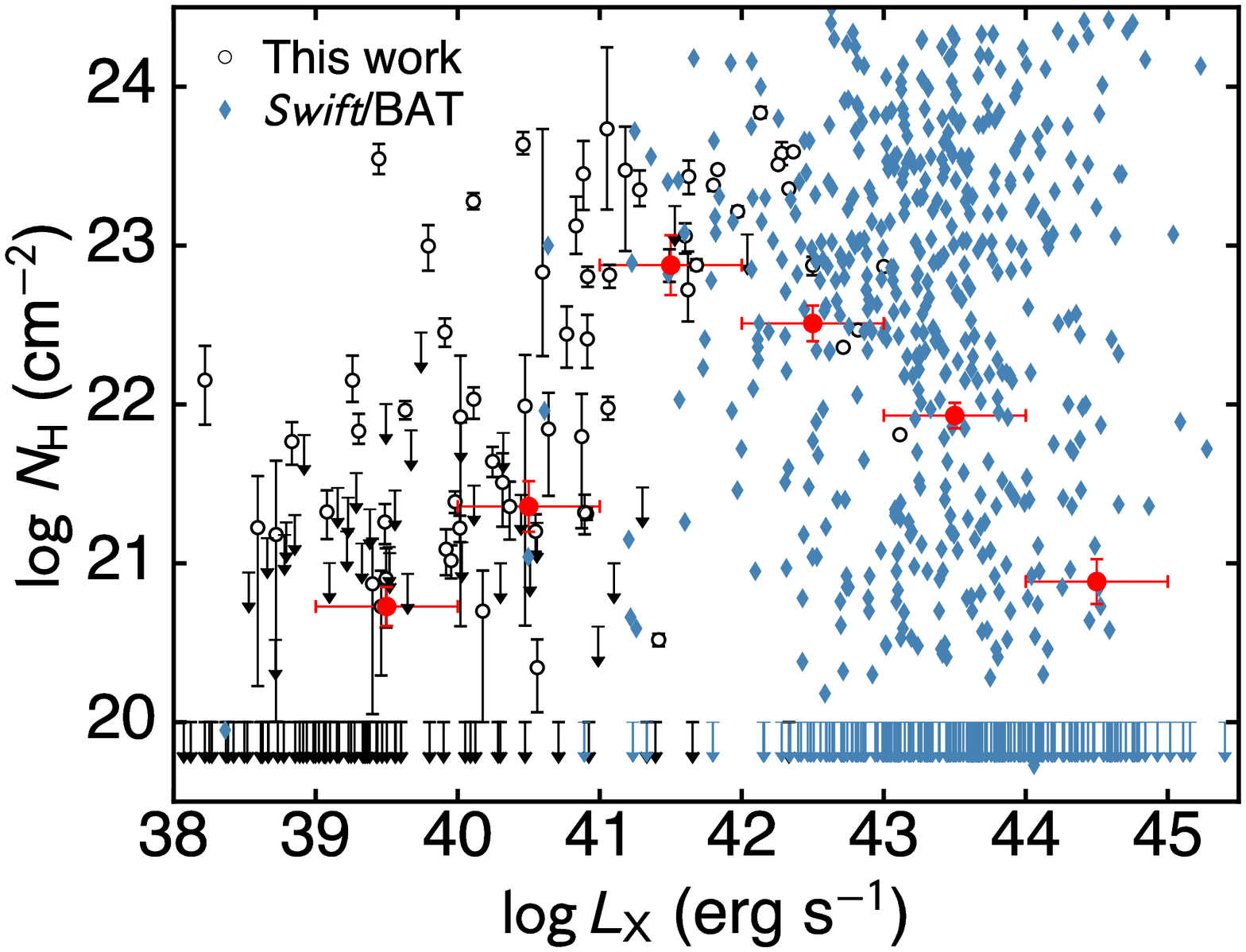}  
\caption{Intrinsic absorption \nhint\ as a function of (top) the Eddington ratio  \eddr\ and (bottom) X-ray luminosity. The red points are the median values with the standard error of the mean (or upper limit) for each bin. The blue data points are adopted from a {\it Swift}/BAT AGN sample \citep{Ricci2017}. The solid lines represent the outflow hydrogen column density at different inclination angles derived from numerical simulations \citep{yuan15}. \label{fig:nh_vs_edd}}
\end{figure}

\begin{figure}[htbp]
\centering
\includegraphics[width=0.77\columnwidth]{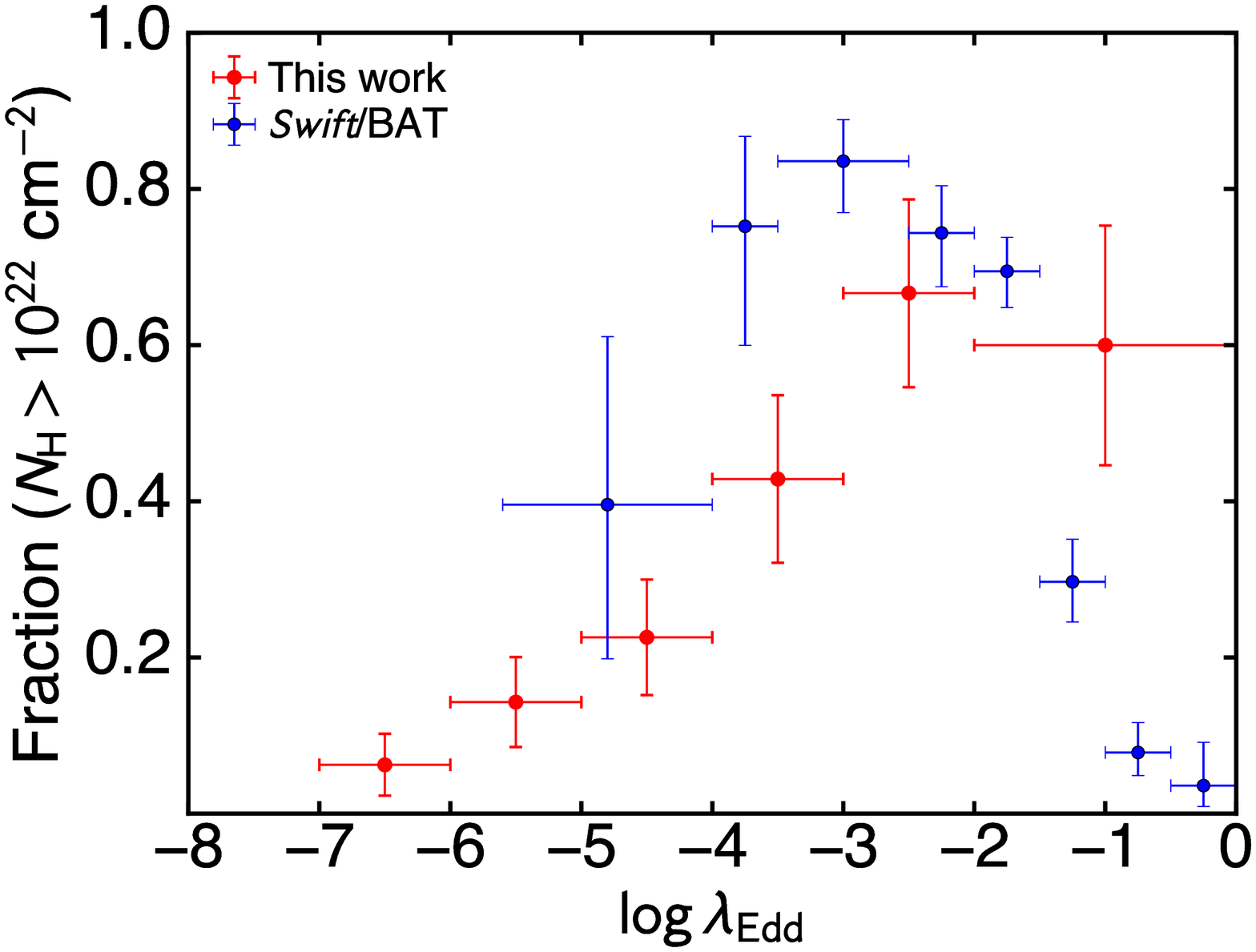} \\
\includegraphics[width=0.77\columnwidth]{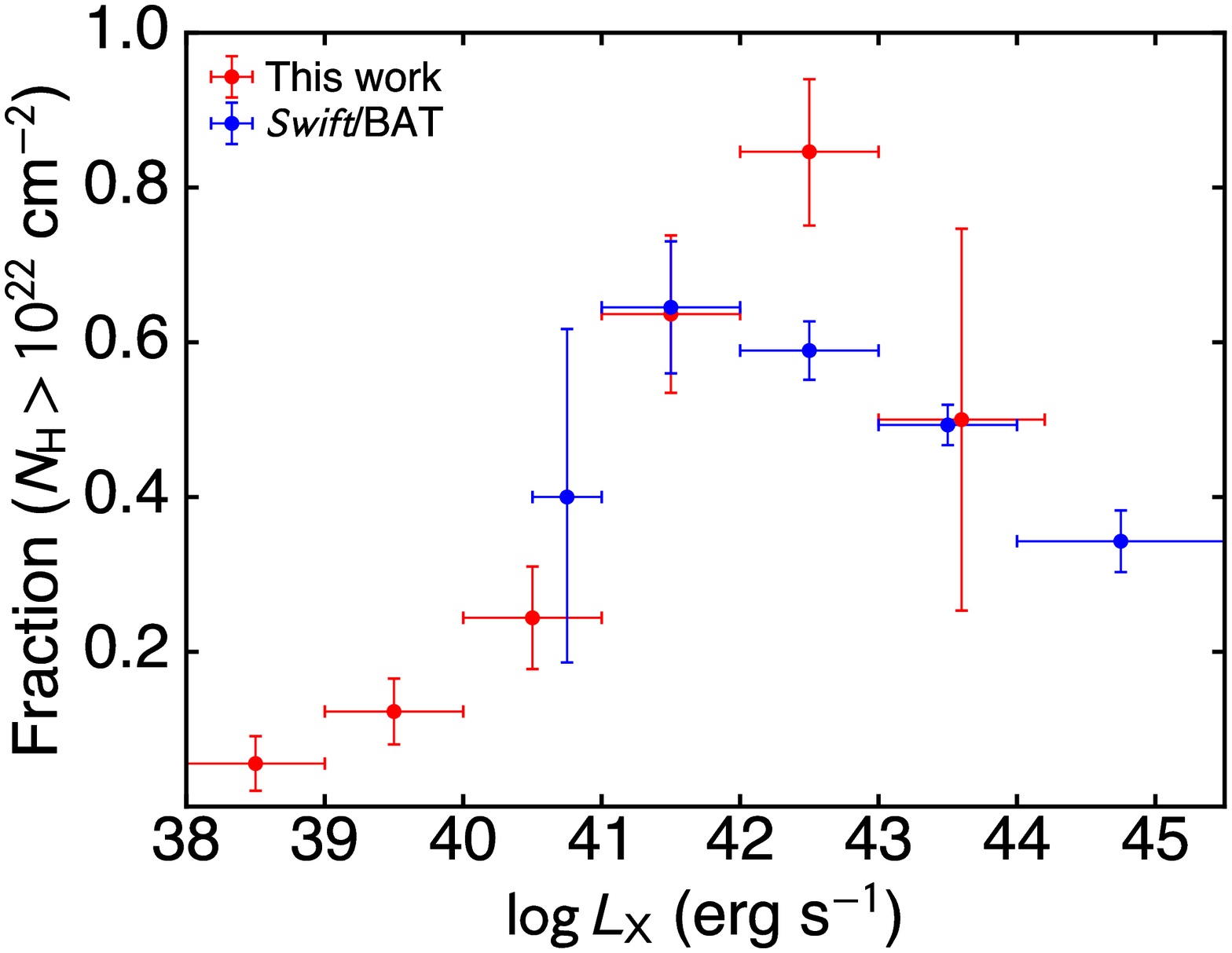} 
\caption{Fraction of absorbed AGNs (\nhint\  $> 10^{22}$~cm$^{-2}$) as a function of (top) the Eddington ratio  \eddr\ and (bottom) X-ray luminosity. \label{fig:frac_vs_edd}}
\end{figure}

The variation of the column density of the intrinsic absorption with X-ray luminosity and Eddington ratio is shown in Figure~\ref{fig:nh_vs_edd}, and the dependence of the fraction of absorbed AGNs (\nhint $> 10^{22} \; {\rm cm}^{-2}$) on X-ray luminosity and Eddington ratio is shown in Figure~\ref{fig:frac_vs_edd}.  As most AGNs in our sample occupy the low-luminosity regime, we add the {\it Swift}/BAT AGN sample \citep{Ricci2017} to expand the dynamic range in luminosity. That sample contains 729 non-blazar AGNs, among which 429 have a black hole mass estimate; the 2--10 keV X-ray luminosities are derived from {\it Swift}/XRT or {\it XMM-Newton} observations. This expands the luminosity range from $10^{36}$ to $10^{46}$~\ergs, or the Eddington ratio from $\sim$$10^{-8}$ to nearly unity.

To compare with numerical results, we also plot the outflow column density versus the Eddington ratio in Figure~\ref{fig:nh_vs_edd}. The solid lines are calculated based on numerical simulation data from three-dimensional general relativistic magneto-hydrodynamic (3D GRMHD) simulations of a hot accretion flow around a supermassive black hole \citep{yuan15}, performed with the HARM code \citep[for details see][]{narayan12}. The simulations indicate that the accretion flow of LLAGNs is hot and covered by an outflow, resulting in a decrease of the local accretion rate with decreasing radius. As there is no radiation included, the simulation is scale-free in density. We calibrate the density of the simulation data at a given luminosity by requiring that the corresponding mass accretion rate produces this value of luminosity according to the radiative efficiency formula given in \citet[][assuming $\delta=0.1$]{xie12}. Note that there may be a systematic error of a factor of a few because the calculation in \citet{xie12} is based on a simple one-dimensional analytical model. The total luminosity was inferred from the innermost accretion rate, following the fitting formula presented in \citet{xie12}. The outflow rate at each radius is found to be proportional to the innermost accretion rate \citep[Equation 5 in][]{yuan15}. Thus, from the simulation results, one can obtain the total column density along the line-of-sight by integrating the outflow from $20 R_{\rm g}$ to infinity, at a given inclination angle (Figure~\ref{fig:nh_vs_edd}).

\begin{figure}[t]
  \centering
  \includegraphics[width=0.77\columnwidth]{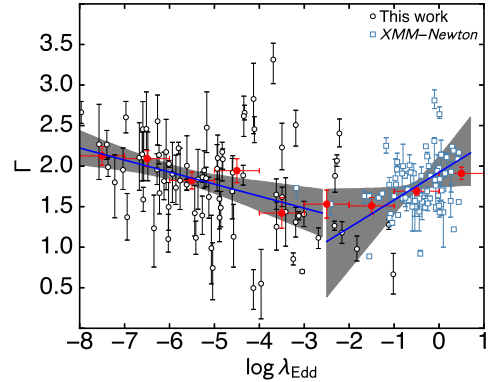}
  \caption{$\Gamma$ as a function of the Eddington ratio \eddr. The blue points are obtained from an {\it XMM-Newton} catalog \citep[CAIXA;][]{bianchi09}. The solid blue lines are the best-fit relations in the low-luminosity (Equation~\ref{equ:gamma1}) and high-luminosity (Equation~\ref{equ:gamma2}) regimes, respectively. The shaded regions indicate the 90\% confidence intervals of the fitted relations. \label{fig:gamma_vs_edd}}
\end{figure}

Figure~\ref{fig:gamma_vs_edd} illustrates the variation of the photon index $\Gamma$ for the power-law component versus the X-ray luminosity or the Eddington ratio. We only include spectra that have a degree of freedom (the number of spectral channels after binning minus the number of free parameters in the model) greater than 5 in the fits, and discard results that have a large error ($> 0.5$). Similarly as above, we add an {\it XMM-Newton} sample \citep[CAIXA;][]{bianchi09} to help increase the statistics on the high-luminosity end.  The X-ray bolometric correction used for our sample is $C_{\rm X} = L_{\rm bol} / L_{\rm X} = 16$ \citep[see Paper I and][]{ho08}. For the CAIXA sample, to be consistent with \citet{ho08}, we adopt $C_{\rm X} = 28$ and 83, respectively, for luminous AGNs and quasars. After including the CAIXA sample (77 objects with a black hole mass estimate), the range of Eddington ratio \eddr\ is $\sim$$10^{-8}$ to $\sim$1. 

\begin{figure*}[tb]
  \centering
  \includegraphics[width=0.35\textwidth]{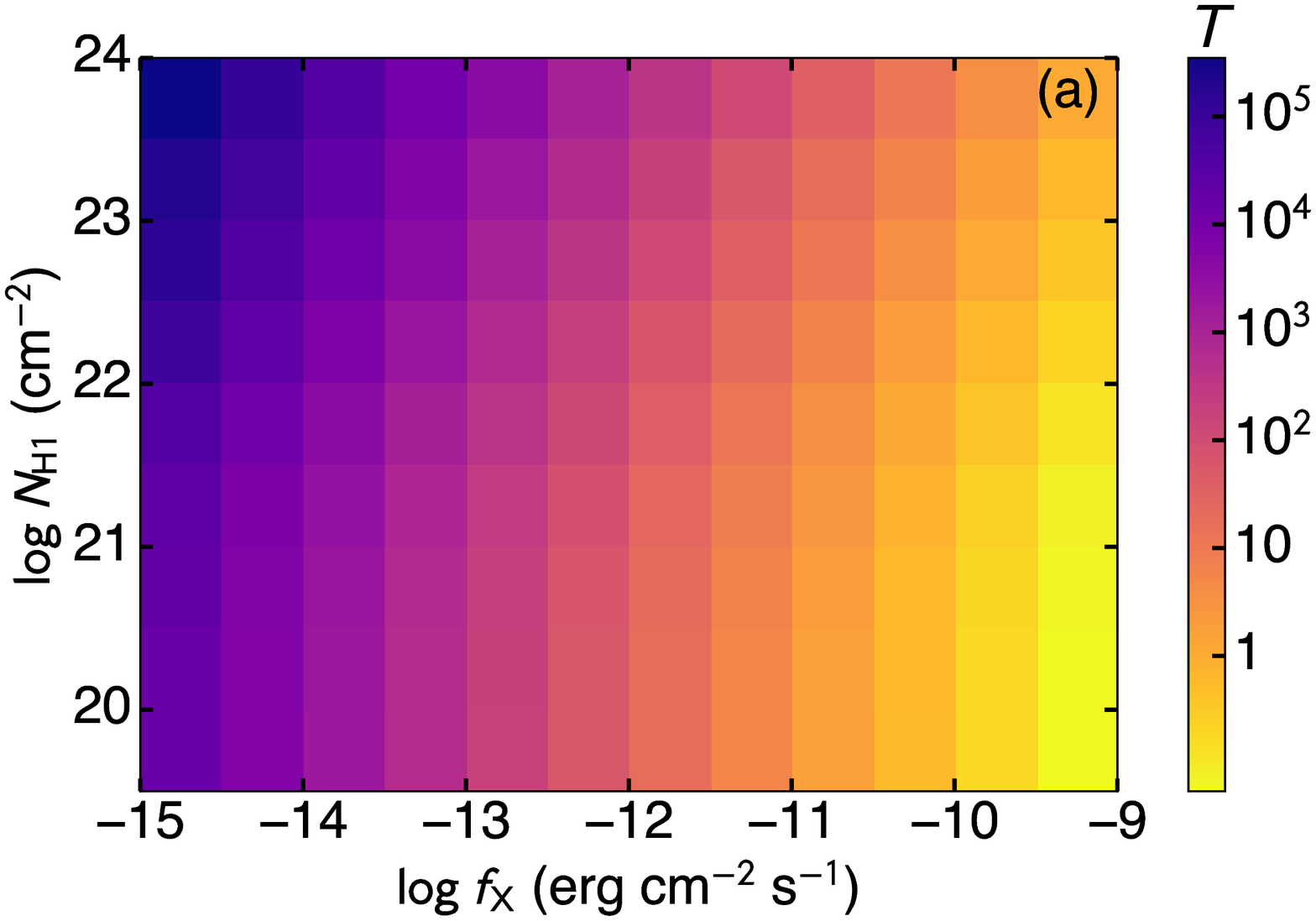}
  \includegraphics[width=0.35\textwidth]{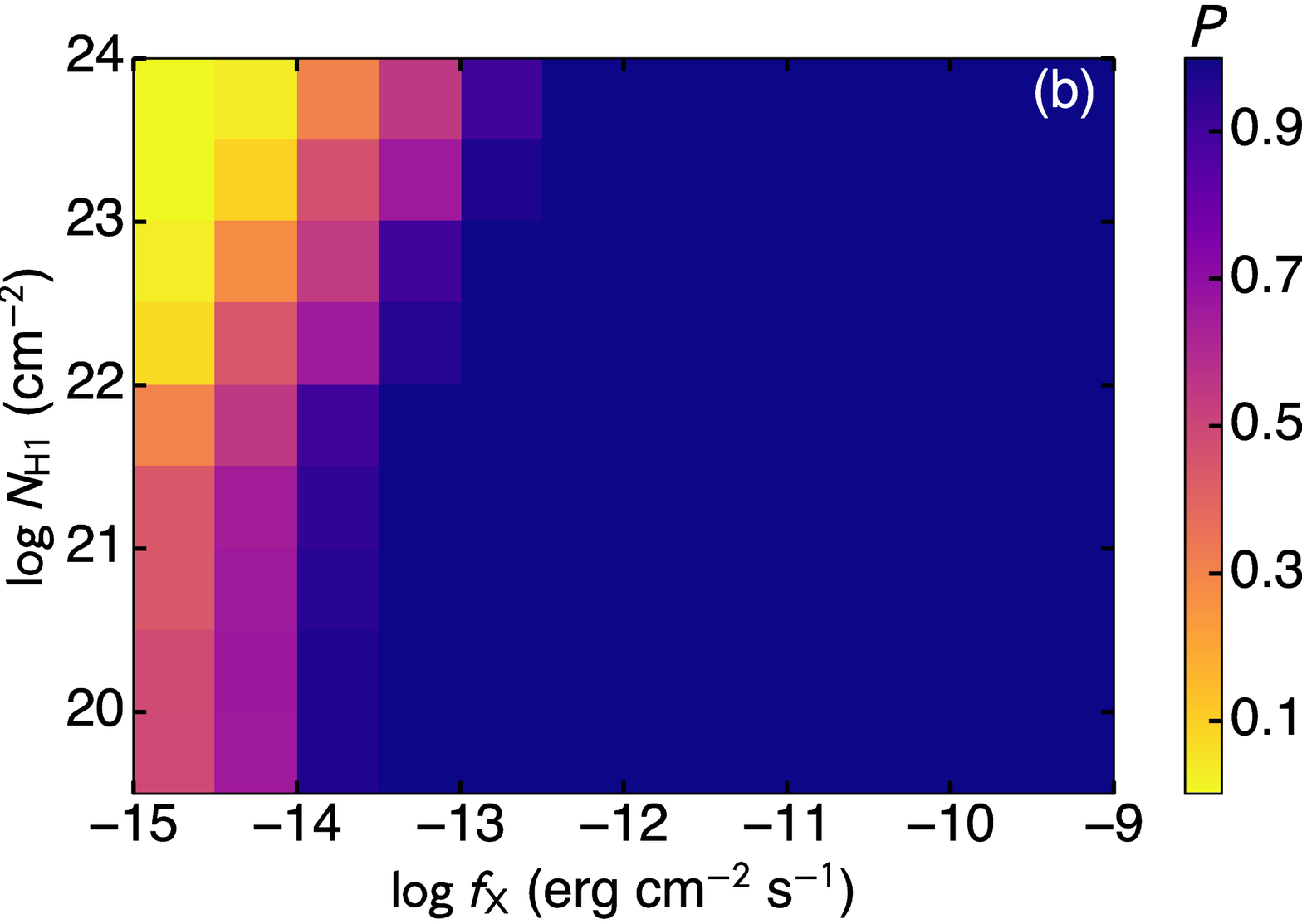} \\
  \includegraphics[width=0.35\textwidth]{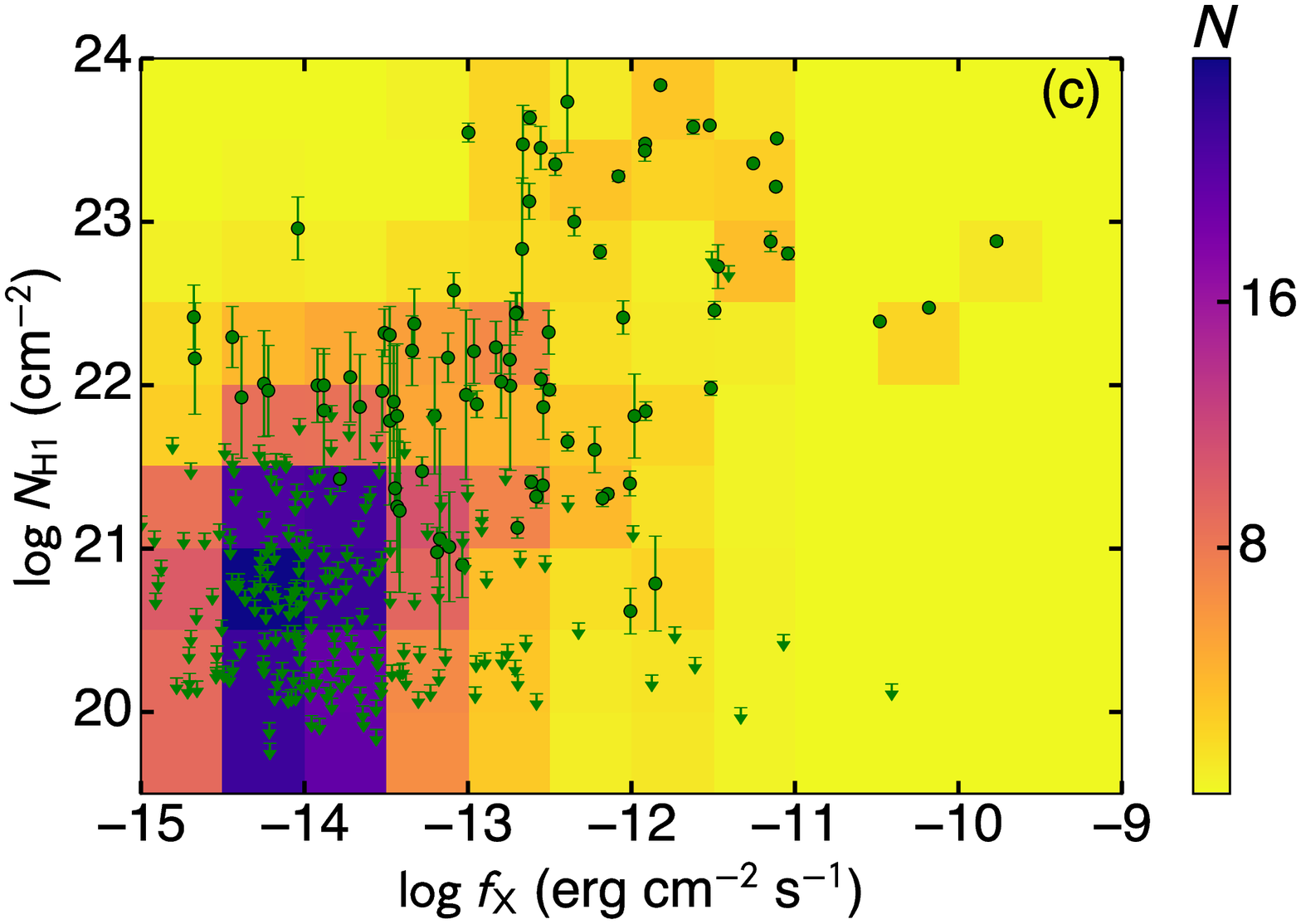}
  \includegraphics[width=0.35\textwidth]{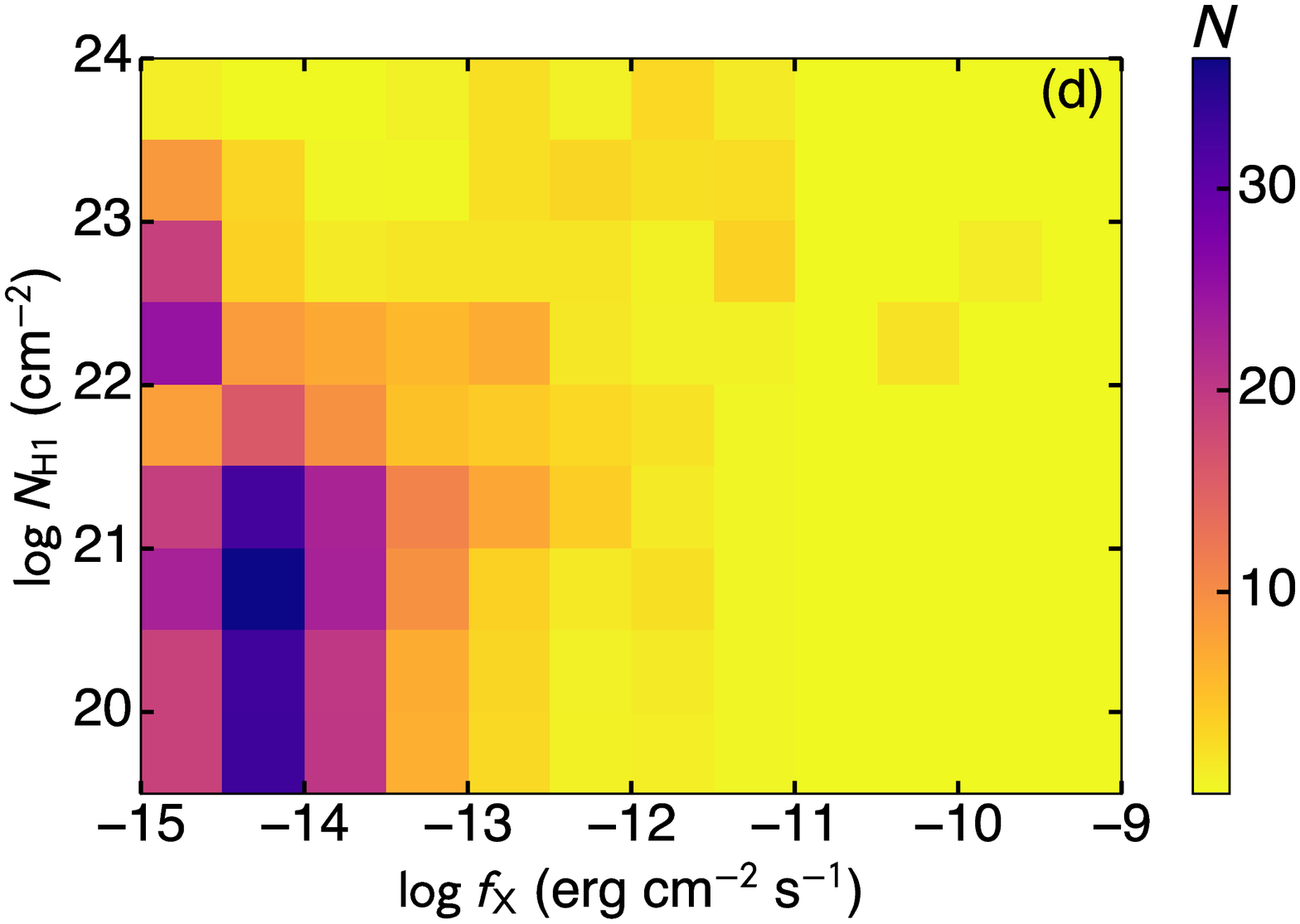}
  \caption{Correction for the observational effect. (a) Exposure times in seconds needed for detecting a source with an absorbed power-law spectrum ($\Gamma=1.8$) given the absorption column density and the 0.3--8 keV X-ray flux. (b) The probability of detecting the source if it is observed with any of the archival \chandra\ observations used by us. (c) The number of detections along with the data points. (d) The intrinsic number distribution after correcting for the detection probability. 
  \label{fig:sim}}
\end{figure*}

We perform a linear regression of the ($y|x$) type to the $\Gamma - \log\lambda_{\rm Edd}$ relation using the bivariate correlated errors and intrinsic scatter (BCES) method \citep{akritas96}, which takes into account both intrinsic and heteroscedastic measurement errors on both $x$ and $y$. The measurement errors on \eddr\ for our sample are not shown in Figure~\ref{fig:gamma_vs_edd}; they mainly arise from the uncertainty on the black hole mass estimate, which has an intrinsic scatter of 0.44 dex (see Paper I).  For AGNs in CAIXA, the black hole mass uncertainty is quoted as 0.46 dex for those with reverberation mapping measurements, or 0.6 dex for those with single-epoch virial estimate \citep{Vestergaard_Peterson_2006}. We split the $\Gamma - \log\lambda_{\rm Edd}$ relation into two branches. For the low-luminosity branch, where $\log\lambda_{\rm Edd} \le -2.5$, 
\begin{equation}
  \Gamma = -(0.15 \pm 0.05) \log \lambda_{\rm Edd} + (1.0 \pm 0.3).
  \label{equ:gamma1}
\end{equation}
For the high-luminosity branch, where $\log\lambda_{\rm Edd} > -2.5$, 

\begin{equation}
  \Gamma = ~(0.3 \pm 0.2) \log \lambda_{\rm Edd} + (1.90 \pm 0.11).
  \label{equ:gamma2}
\end{equation}
The intrinsic scatter is 0.3 for both branches. The regression line and 90\% confidence intervals of the relation are shown in Figure~\ref{fig:gamma_vs_edd}. 

From an observational perspective, the {\tt mekal} component is more frequently detected in less luminous sources. We further test the $N_{\rm H}$-$\lambda_{\rm Edd}$ and $\Gamma$-$\lambda_{\rm Edd}$ relations for objects without a {\tt mekal} component.  We find consistent results, suggesting that the treatment of the diffuse hot gas background and inclusion of the {\tt mekal} component do not affect our conclusions.

\section{Testing for observational effects}

The same trend seen between \nhint\  and $\lambda_{\rm Edd}$ or $L_{\rm X}$ (Figure~\ref{fig:nh_vs_edd}) is also evident if we plot the absorption column density against the observed flux $f_{\rm X}$ (Figure~\ref{fig:sim}c); that is, the sources seem to be less absorbed at lower flux.  Such a phenomenon may be produced by an observational effect, as faint objects with high absorption are hard to be detected. However, the \nhint\  versus $L_{\rm X}$ correlation may be real. Because the dynamical range of the distance for these objects (within a factor of 10 for 90\% of the sources) is much smaller than that of the luminosity ($\sim$4 orders of magnitude), a correlation with luminosity may also lead to a correlation with flux. 

It is the total absorption \nhtot\ that affects the detection. The Galactic \nhgal\ is usually much smaller than the intrinsic \nhint\ if it is detected. The correlation seen in Figure~\ref{fig:nh_vs_edd} remains if we plot it against \nhtot\ instead. We thus try to remove the observational effect to see if the correlation between \nhtot\ and $f_{\rm X}$ remains. We divide the $\log$ \nhtot\  $- \log f_{\rm X}$ space into small pixels. For each pixel, the central \nhtot\ and $f_{\rm X}$ is selected to simulate a spectrum assuming an absorbed power-law spectrum with $\Gamma = 1.8$. Then, the exposure time needed to detect such a source with at least 7 counts with {\it Chandra} ACIS is calculated (Figure~\ref{fig:sim}a). Assuming that we randomly choose a \chandra\ observation to observe this target, the probability of detection can be inferred from the distribution of the exposures for all of the \chandra\ observations in the archive used (observations for the \GalNum\ galaxies), which is defined as the map of detection probability (Figure~\ref{fig:sim}b). Given the measured \nhtot and $f_{\rm X}$ and their errors, we fill the pixels in the $\log$ \nhtot\  $- \log f_{\rm X}$ space using a two-dimensional Gaussian kernel. In some cases, when the lower error is not available, all pixels below the center are filled with a uniform distribution summed to 0.5. For objects with a non-detectable intrinsic \nhint\  ($< 10^{20}$~cm$^{-2}$),  the pixel (usually the very bottom one) that contains $N_{\rm H0}$ is filled with unity.  This converts discrete detections into a map of detection numbers (Figure~\ref{fig:sim}c) . The division of the detection number by the detection probability reveals the intrinsic distribution of objects on the  $\log$ \nhtot\ versus $\log f_{\rm X}$ plane after correction for the observational effect.  As one can see, a positive correlation remains in the central region of the plot. Also, at the smallest \nhtot, the number of non-detections increases toward low fluxes, which strengthens the argument. The high-density region at the smallest fluxes is due to only two data points and is thus associated with large uncertainty. Therefore, we argue that the positive correlation between \nhtot\ (and thus \nhint)  and $f_{\rm X}$ is not an observational effect; instead, it is a result of a genuine correlation between the absorption and luminosity. A detection with 7 counts is assumed for on-axis ACIS observations; for observations with a reasonable off-axis angle, the number will become slightly larger but the conclusion remains. 

\section{Discussions}
\label{sec:discussion}

Using a nearby AGN sample constructed in Paper I, we show here that the intrinsic absorption of LLAGNs positively scales with increasing Eddington ratio, and that the X-ray spectrum becomes softer with decreasing accretion rate. Both relations are seen at $\lambda_{\rm Edd} < 10^{-3}$, opposite from the source behavior at higher accretion rates, indicative of a change of the accretion physics at low accretion levels. 

A negative correlation between the level of X-ray absorption and luminosity at the high-luminosity branch has been reported extensively in the literature \citep{ueda03,lafranca05,akylas06,burlon11}. Recently, \citet{Ricci2017} found that the probability that a source is obscured in the X-rays (covering factor of dusty gas) depends primarily on Eddington ratio instead of on absolute luminosity.  \citet{Ricci2017} propose that the radiation pressure on dusty gas is responsible for regulating the distribution of obscuring material around the central black hole.  At high accretion rates, radiation pressure expels the obscuring material in the form of outflows \citep{fabian06}.

In the low-luminosity regime, this is the first time that a relation between the intrinsic absorption and the Eddington ratio is reported for a large sample.  Similar results in previous studies \citep[e.g.,][]{beckmann09,zhang09} either were based on a small sample or were discussed in the context of absolute luminosity instead of the Eddington ratio.  \citet[their Figure 1]{Ricci2017} also find a tentative decrease of the fraction of obscured sources for log~$\lambda_{\rm Edd} \lesssim -4$, but the statistical significance is low because of the small number of LLAGNs in their sample.  A consistent relation between the outflow column density and  Eddington ratio derived from the numerical simulations of \citet{yuan15} suggests that the X-ray absorption in LLAGNs may originate from or is related to outflowing material. We argue that the column density from the simulation can be regarded as an upper limit, as radiation transfer (e.g., scattering) in the outflow may help increase the observed luminosity, particularly in the soft band.  Future simulations should take into account the radiation transfer, ionization, and metallicity to address this problem more accurately. 

We do not know whether or not the absorption arises from gas and dust on larger scales,  which may influence both the accretion rate and absorption column density. Our current data cannot disentangle the two models, but in the future it may be useful to perform a correlation analysis between the accretion and absorption on timescales close to the accretion timescale but much smaller than that for the large-scale structure.

The $\Gamma$--\eddr\ relation in the high-luminosity regime can be explained if the X-rays arise from a corona due to Comptonization of soft photons from the disk: higher accretion rate leads to enhanced seed photons or a larger cooling power, and consequently a softer spectrum \citep{chakrabarti95,zdziarski99}. The inverse correlation between  $\Gamma$ and \eddr\ in the low-luminosity regime  is consistent with previous studies \citep{gu09a,younes11}.   At very low accretion rates, the hot flow is responsible for both the seed photons, via synchrotron or bremsstrahlung radiation, and the Comptonization \citep{yuan14a}, and the observed inverse correlation  between  $\Gamma$ and \eddr\ is expected in ADAF models.  \citet{yang15} found that such a relation can be explained within the framework of the ADAF model and the luminous hot accretion flow model, and that the transition around $\lambda_{\rm Edd} \sim 10^{-3}$ may correspond to the transition of the accretion mode.

In our sample, 64 objects are classified as type 2 AGNs in the optical, but most of them have low levels of intrinsic absorption, with a median column density of $1.7 \times 10^{21}$~cm$^{-2}$; 21 of them even do not allow for a positive detection of the intrinsic absorption. The weakness of X-ray absorption in optical type 2 LLAGNs suggests that these sources may intrinsically lack a dust torus and are {\it intrinsic}\ type 2 objects. This is consistent with previous studies and suggests that the broad-line region and dust torus are associated with AGN outflows \citep{elitzur06,elitzur09}, which vanish for sufficiently low accretion rates \citep{ho08}. 

To summarize: in the LLAGN regime, the dependence of the intrinsic X-ray absorption and spectral hardness with accretion rate can be self-consistently explained if the central supermassive black hole possesses a hot accretion flow undergoing significant outflow with a mass loss rate that correlates with the innermost mass accretion rate.  The \nhint--\eddr\ relation can be used as an effective test of future theoretical or numerical simulations of hot accretion flows. 

\acknowledgments
We thank the anonymous referee for useful comments that have helped improve the paper. We thank Claudio Ricci for providing the \textit{Swift} data and useful discussions.  We are grateful to Feng Yuan for valuable comments and sharing results from his simulations.  LCH was supported by the National Key R\&D Program of China (2016YFA0400702) and the National Science Foundation of China (11473002, 11721303).  HF acknowledges funding support from the National Natural Science Foundation of China under grant No.\ 11633003, and  the National Program on Key Research and Development Project (grant No. 2016YFA040080X).  


\end{document}